\documentclass[a4paper]{jpconf}
\usepackage{graphicx}
\usepackage{hyperref}
\usepackage{enumitem}

\newcommand{\subopzero}{0}
\newcommand{\subopone}{1}
\newcommand{\suboptwo}{2}

\begin{document}
\title{A definition  of the SI second based on several optical transitions}

\author{Jérôme Lodewyck}

\address{LNE–SYRTE, Observatoire de Paris, Université PSL, CNRS, Sorbonne Université,
61 avenue de l'Observatoire, 75014 Paris, France}

\ead{jerome.lodewyck@obspm.fr}

\begin{abstract}
A new definition of the SI second based on optical transitions is expected to be adopted within the next 10 years. Several options for this redefinition are currently under consideration. Among them, a definition based on several transitions would take advantage of the variety of high performance optical frequency standards. In this paper, we review practical aspects such a definition entails, and propose a detailed analysis of its strengths and weaknesses.
\end{abstract}

\section{Introduction}

The rapid progress in optical frequency standards over the last two decades led the Consultative Committee for Time and Frequency (CCTF) to craft a roadmap~\cite{Dimarcq_2024} for the future redefinition of the SI (International System of Units) second based on optical clock transitions, aimed at replacing the current definition based on the caesium atom. Three options are under consideration:
\begin{itemize}
	\item \emph{Option 1:} a single atomic transition $i$ replaces to the Cs hyperfine transition. Its frequency would be the new defining constant for the second, and its numerical value would be set to:
	\begin{equation}
		\label{eq:simple}
		\nu_i = N{\textrm{ Hz}}
	\end{equation}
	where the exact numerical constant $N$ is the defining value.
	\item \emph{Option 2:} the definition is based on the weighted geometric mean of the frequencies of several transitions. This average, having the dimension of a frequency, would be the new defining constant for the second, and its numerical value would be set to:
	\begin{equation}
		\label{eq:geom}
		\prod_i \nu_i^{w_i} = N{\textrm{ Hz}}
	\end{equation}
	where the exact normalisation constant $N$ and the weights $w_i$ (summing to one) are the defining values.
	\item \emph{Option 3:} the second is defined by fixing the numerical value of a fundamental constant. For instance, the electron mass $m_e$ would be the new defining constant, and its numerical value would be set to:
	\begin{equation}
		m_e = M{\textrm{ kg}}
	\end{equation}
	where the exact numerical constant $M$ is the defining value.
\end{itemize}

Option 1 is a natural continuation of the current definition of the second, simply replacing caesium with another atomic species. Option 3 is an intellectually appealing extension of the 2018 definitions of the other base units of the SI, with the caveat that the electron mass is a property of a specific particle, unlike the other constants used in the SI, which are more universally rooted in the fundamental equations of physics. However, because the realisation uncertainty of such constants is many orders of magnitude larger than the uncertainty of the best frequency standards, this options is deemed unpracticable.

Option 2 has been introduced in~\cite{Lodewyck_2019}. The motivation for this proposition is two-fold: First, the large number of optical frequency standards with equivalent performances makes it difficult to reach a consensus and to single out one of them as a new Primary Frequency Standard (PFS). Second, in order to cope with the diversity of frequency standards, the Consultative Committee for Length (CCL) and CCTF Working Group on Frequency Standards (WGFS) has put in place a procedure that consists in collecting all measured frequency ratios between these various primary and secondary frequency standards in order to derive, by a least square procedure~\cite{pub.1058980983}, a consistent set of best estimates for these frequency ratios. Although the output of this procedure is formally a set of ``recommended frequencies for Secondary Representations of the Second (SRS)''\cite{Riehle_2018}, hence linked to the Cs transition, the genuine mathematical output of the fit is actually a set of frequency ratios (given by the ratio of recommended frequencies) together with their covariance matrix (with uncertainties that can be smaller than the uncertainties on recommended frequencies). It is  thus completely independent of the definition of the second, and, more generally,  does not distinguish a particular transition. In light of this \emph{de facto} procedure, option 1, as a continuation of the current definition of the second, arguably maintains an artificial, arbitrary, and to some extent unnecessary bias in the definition of the second. Therefore, a more ``decentralized'' definition, based on the fit output would be a more elegant choice, practically taking advantage of the many frequency ratio measurements currently available. Such an idea was already put forward by P. Gill at the 8th Symposium on Frequency Standards and Metrology~\cite{Gill_2016}:
\begin{quote}
 [\ldots] a definition comprising a ``universal'' value determined from the matrix of individual best values for different species.
\end{quote}
In~\cite{Lodewyck_2019} we explicitely proposed a physical quantity with the dimension of a frequency, namely the geometric mean~(\ref{eq:geom}), derived from the ``matrix'', that could effectively give corpse to this ``universal'' value. The discussion in~\cite{Gill_2016} continues with:
\begin{quote}
However, in this scenario, the definition would not be based on a fundamental constant or a ``constant of nature'' (as in the current definition), but on an aggregation of values for different systems, with little physical significance. Further, this value would change every time individual values were upgraded, and change significantly with the introduction of a new species into the value matrix.
\end{quote}
This notice illustrates the questions and concerns raised by a definition that does not rely on a single atomic transition. In this paper, we aim at clarifying practical aspects of option 2 (section~\ref{sec:practical}), and at proposing a strengths and weaknesses analysis answering potential concerns about option 2, such as its physical significance as cited above (section~\ref{sec:proandcons}). Finally, we discuss the possibility and the need for regular and smooths updates of the defining values~\cite{Lodewyck_2019}  (section~\ref{sec:updates}).

\section{Practical implementation of option 2}
\label{sec:practical}

\subsection{Realizing the unit}
For a definition to be practical, it must be possible to practically realize it. In option 2, the definition of the SI second~(\ref{eq:geom}) could in principle be directly realised by building a complete set of clocks, each realizing one of the transitions entering the definition, and by combining their output. More precisely, measuring the frequency $\nu_\textrm{\tiny osc}$ of an oscillator\footnote{For instance, the local pivot oscillator used to steer the International Atomic Time (TAI)} in Hz would involve measuring its frequency ratio $\rho_{\textrm{\tiny osc}, i}$ against all clocks $i$ and then deduce, knowing the exact numerical constants $N$ and $w_i$:
\begin{equation}
	\label{eq:nuoscall}
	\nu_\textrm{\tiny osc} = N \prod_i \rho_{\textrm{\tiny osc}, i}^{w_i}\ \textrm{Hz}.
\end{equation}

\begin{figure}
	\begin{center}
		\includegraphics[width=\textwidth]{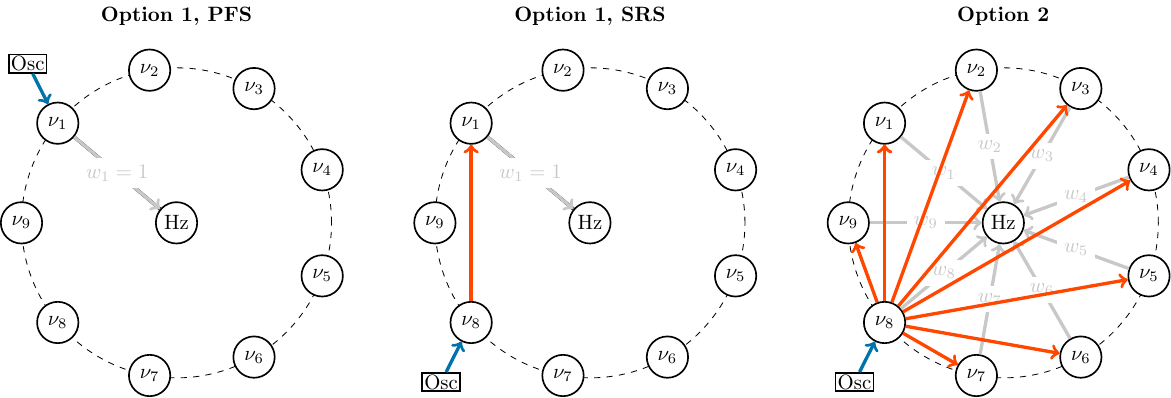}
	\end{center}
	\caption{\label{fig:connection}Realizing the second in options 1 and 2. Option 1 can be seen as a specific case of option 2 for which a single transition (here labelled 1) has a non-zero weight: $w_i = \delta_{i,1}$. Measuring the frequency of an oscillator in the SI can be done either by directly comparing this oscillator to a Primary Frequency Standard (left figure), or to a Secondary Representation of the Second (SRS) (middle figure) knowing the recommended frequency for this SRS (orange arrow). In option 2, the oscillator can be equivalently compared to any frequency standard entering the definition, using its recommended frequency. It is connected to the unit through the global fit of frequency ratios between the various frequency standards.}
\end{figure}

In practice, building a full set of clocks realizing all the transitions composing the unit can be involving. But even if only a single clock $i_0$ is available, it is still possible to determine $\nu_\textrm{\tiny osc}$ in Hz, by relying on otherwise known frequency ratios $\rho_{i,j}$ between the clock transitions:
\begin{equation}
	\label{eq:nuosc1}
	\nu_\textrm{\tiny osc} = \rho_{\textrm{\tiny osc}, i_0} N_{i_0}\ \textrm{Hz} \quad  \textrm{with} \quad N_{i_0} = N \prod_i \rho_{i_0, i}^{w_i}.
\end{equation}
The numerical constant $N_{i_0}$ is independent of the oscillator being measured, and can be viewed as the recommended frequency for transition $i_0$, the equivalent of the recommended frequency for SRS in the current definition of the second (see figure~\ref{fig:connection}).


The output of the global fit of frequency ratios by the WGFS can be used to compute the recommended frequencies $N_i$, and they can be regularly published and updated by the CIPM, in the same way as the recommended frequencies of SRS are currently published. The covariance matrix output by the global fit can also be used to provide uncertainties for these recommended frequencies (see equation (A.5) of reference~\cite{Lodewyck_2019}), which add up to the realisation uncertainty of the frequency standard used to realized the unit. This added realisation uncertainty fundamentally comes from our limited knowledge of frequency ratios.

Equations~(\ref{eq:nuoscall}) and~(\ref{eq:nuosc1}) correspond to the extreme cases for which all transitions and a single transition, respectively, are used to realise the unit. They can be easily extended to any intermediate cases, and feature an added realisation uncertainty that decreases as more clocks are implemented.

\subsection{Choice of defining values}

The unit is defined by its defining values $N$ and $w_i$. In this section, we detail the procedure to determine these values, set at the time of the redefinition.

\subsubsection{Weights}

In order to reach a consensus on the implementation of option 2, it is preferable to choose the weights $w_i$ of the various species composing the unit following systematic and quantitative rules. In~\cite{Lodewyck_2019}, it was shown that, in the special case of a unit composed of two transitions, the ratio $w_2/w_1$ should be strictly larger than the inverse ratio between the clock uncertainties $u_1/u_2$, so that the realisation uncertainty of the unit with the best clock would not be significantly hindered by the uncertainty of the other clock. From this result, it sounds reasonable to choose weights proportional to the inverse \emph{squared} uncertainty $1/u_i^2$. More generally, it can be proven by a least square calculation that, for an arbitrary number of clock transitions, this choice actually minimizes the overall added realisation uncertainty.

The choice of weights proposed above requires that the clock uncertainty $u_i$ is a well-defined quantity. However, the uncertainty $u_i$ is different from realisation to realisation, and is thus not intrinsic to the clock transition itself. Therefore, a method to estimate a typical uncertainty $u_i$ for the transition $i$ must be established, using the set of reported realisation uncertainties. Because the ultimate aim of a unit is to effectively be able to build frequency standards capable of operational measurements, it seems reasonable to select reported uncertainties of clocks that were involved in actual high precision frequency ratio measurements. In addition, it is preferable that a transition that is connected to other transitions only by a single frequency ratio measurement should not have a too large weight in the unit, to ensure redundancy in the evaluation of frequency ratios. From these considerations, a heuristic method was proposed in~\cite{Lodewyck_2019}: the weight $w_i$ is obtained by averaging the inverse squared relative uncertainties of the two best frequency ratios involving the transition $i$. Although this method, and variations thereof, was found to provide reasonable and consistent weights when applied to the set of frequency ratios published before the 2017 CCTF, it lacks rigorous foundations, and also prevents option 2 from eventually converging to option 1: if a single transition stands out, its weight will be at most identical to the weight of the second best transition. To resolve these issues, we developed a rigorous method to set the weights, in the spirit of the three-cornered hat method: $u_i$ is obtained as the output of a least squares adjustment that aims at matching the relative uncertainty on the frequency ratio $\rho_{i,j}$ with the quadratic sum of $u_i$ and $u_j$. This method takes as input the covariance matrix of the outcome of the global fit of frequency ratios: by construction, it only relies on uncertainties associated with frequency ratio measurements. Furthermore, it provides a fair estimate of the uncertainty of a single clock transition, provided a realisation of this transition have been compared to at least two other transitions.

An independent publication in preparation will aim at detailing the results of this section.

\subsubsection{Normalization constant}

The normalisation constant $N$ is a fixed, exactly known constant, that is chosen to ensure the continuity of the unit across the redefinition. In option 2, as well as in option 1, this choice is driven by the set of frequency ratio measurements available at the time of the redefinition: $N$ is computed from the set of recommended frequencies derived by fitting the set of frequency ratios. Quantitatively, the normalisation constant is chosen such that the recommended frequency for the caesium clock transition at the moment of the redefinition is identical to the previously fixed value. The uncertainty on this recommended frequency yields the typical, unavailable mismatch between the new and the old definition. Such a mismatch, arising from our limited knowledge of frequency ratios is unavailable, and on the same order, for both options 1 and 2.

\subsection{After the redefinition}

After the redefinition, the various optical frequency standards will likely be further improved, and new and more precise frequency ratios between theses standards will be measured. We can thus legitimately ask whether these improvements may \emph{change} the definition of the second, or worse, be \emph{incompatible} with the definition. After all, would it be possible that updated measured frequencies for the various transitions composing the unit would eventually not geometrically average any-more to the fixed value chosen for the definition? This concern can be answered with two properties of option 2:
\begin{enumerate}
	\item \emph{The weights $w_i$ and the normalisation constant $N$ univocally define the unit}, without the need to specify a set of measured frequency ratios. Indeed, if we were to specify the duration of a second to an extra-terrestrial civilization, the sole transmission of $w_i$ and $N$ would be sufficient. From these values, the alien civilization would be able to realize the second, by simply building clocks implementing each species composing the unit and applying equation~(\ref{eq:nuoscall}), possibly with a lower realisation uncertainty that we can currently attain ourselves. The limited knowledge of frequency ratios only limits our ability to \emph{realize} the otherwise completely specified definition of the second, and acquiring more precise values for these ratios can only help reducing the realisation uncertainty, in the same way as acquiring a more precise value of atomic polarisability or collision cross-section can help reducing the realisation uncertainty of the current definition SI second based on caesium. Said otherwise, the knowledge of physical quantities (frequency ratios for Option 2, polarisability, cross-sections,\ldots for Options 1 and 2) is necessary to connect a physical realisation of a frequency standard, to the actual, idealized, definition of the unit; and uncertainties on these quantities constitute the realisation uncertainty of the perfectly defined unit.
	\item \emph{The normalisation constant $N$ is unconstrained by frequency ratio measurements}. Assuming new clock comparisons yield refined measured frequency ratios after the redefinition, with improved uncertainties, these new measurements will allow us to derive improved recommended frequencies for all the transitions composing the unit. But while the recommended frequencies do change with improved frequency ratios measurements, their weighted geometric mean stays invariably exactly equal to $N$, regardless of the actual values of measured frequency ratios.

	In order to prove this fact, we note $\bar \rho_{i,j}$ the set of best estimates for frequency ratios as output by the least squared global fit of frequency ratio measurements. By construction, they satisfy the relations $\bar \rho_{i,i} = 1$ and $\bar \rho_{i,j}\bar \rho_{j,k} = \bar \rho_{i,k}$, which are constraints built in the fit. The updated recommended frequencies would then read $N_i = N\prod_k \bar\rho_{i,k}$, from where it can be shown that $\prod_i N_i^{w_i} = N$, independently of the actual values of $\bar \rho_{i,j}$:
	\begin{equation}
		\prod_i N_i^{w_i} = \prod_iN^{w_i}\prod_k \bar\rho_{i,k}^{w_iw_k} = N \frac{\prod_{i,k}\bar \rho_{i,i_0}^{w_iw_k}}{\prod_{i,k}\bar \rho_{k,i_0}^{w_kw_i}} = N \frac{\prod_{i}\bar \rho_{i,i_0}^{w_i}}{\prod_{k}\bar \rho_{k,i_0}^{w_k}} = N,
	\end{equation}
	where we mathematically used a specific, arbitrary transition $i_0$ as a pivot.
\end{enumerate}

\section{Comparison between options 1 and 2}
\label{sec:proandcons}

Choosing between the various options for the redefinition of the second requires to compare the advantages and drawbacks of each option. In this section, we propose a detailed analysis the strengths and weakness of option 2 that have been identified in the roadmap~\cite{Dimarcq_2024}, focusing on those for which option 2 differs from option 1.

\subsection{Strengths}

Option 2 brings several advantages, as compared to option 1:

\begin{itemize}
	\item \emph{Reaching a consensus}. There are currently many possible technologies (ion clocks and optical lattice clocks) and atomic species for the redefinition of the SI second, with comparable performances in the low $10^{-18}$. With option 1, a unique atomic species is singled out, introducing a somehow arbitrary distinction between the possible candidates. Reaching a consensus on a single species can potentially be a difficult task, and may delay the redefinition of the second. Option 2 solves this choice problem by attributing weights to the different candidates, objectively based on their demonstrated performances in measuring high accuracy frequency ratios. Option 2 would thus lead to an easier and earlier consensus.
	\item \emph{Promoting diversity}. While the vast variety of optical frequency standard is a challenge for the redefinition of the second, it is a wealth for the scientific community, because it encourages innovation, and large data sets of interspecies comparisons are helpful to improve the robustness of optical frequency standards, and to test fundamental physics. Option 1 may lead to a strong focus of resources on a single species, which will tend to lessen the applications of optical frequency standards. On the contrary, option 2 not only does not strongly bias the definition on a single species dubbed ``primary frequency standard'', but also specifically encourages the production of new and more precise frequency ratio measurements, in order to decrease the realisation uncertainty. Said otherwise, the best way to improve the realisation of the SI second with option 1 is to build several PFS of the same kind, while in option 2, it is to build several different frequency standards and to compare them.
	\item \emph{Readily and simply implementable}. All the tools to implement option 2 are already in place: the choice of defining values and the publication of recommended frequencies directly rely on the global fit of frequency ratios, for which the WGFS has now a decade-long experience. Steering the International Atomic Time (TAI) will be done in the same was as it is currently done with SRS.
	\item \emph{Mitigate risks}. Despite the best effort of the time and frequency community in implementing frequency standards, it remains possible that a particular systematic bias has been overlooked, or wrongly estimated for a specific clock transition. Such a mistake would have a different impact on the realisation of the second in option 1 and 2: on the one hand, it would only impact option 1 if it affects the transition chosen as the definition, whereas it would impact option 2 if it affects any transition entering the definition. On the other hand, the error would be reduced in option 2 because the transition only partly contributes to the definition. Likewise, if, at the redefinition, a recommended frequency happens to have an underestimated uncertainty, the step at the redefinition would be more probable with option 2, but smaller. In summary, option 1 favours high risk low probably events, whereas option 2 favours lower risk but higher probability events. A more far-fetched application of option 2 is the possibility to choose the weights of transitions in order to make the unit insensitive to a putative variation of $\alpha$.
\end{itemize}

\subsection{Weaknesses}

Nevertheless, option 2 is a rather disruptive way to define a unit, and hence triggers questions and concerns. Here, we give a list of possible drawbacks of option 2, together with arguments mitigating them.

\begin{itemize}
	\item \emph{Match with the spirit of the SI}. With the introduction of a fixed value for the speed of light in 1983, and a fixed value for other physical constants in 2018, the spirit of the SI system is to base units on the laws of physics. Option 2 contrasts with this trend, with a definition that is more driven by technological readiness than fundamental considerations. However, it can be argued that the aim of a system of unit should first to be practical, and then, among all practical options, the most elegant and fundamental one can be selected. For instance, it was only possible to redefine the kilogram by fixing the Planck constant after the uncertainty of practical realisations could allow for it. This is the reason why option 3, although more elegant than other options, is not considered as a viable option for the redefinition. In contrast, equation~(\ref{eq:simple}) for option 1, or equation~(\ref{eq:geom}) for option 2, essentially consist in fixing the numerical value of a function of fundamental constants, solution of the physical equations describing the state of one or more many-electron atoms. Although fixing such a function amounts in effect to fixing a combination of fundamental constants, this function is currently unknown at the level of precision of the best realisation of the second, even for an atom as simple as hydrogen. Consequently whether this function involves a single or several atomic species is essentially insignificant from the point of view of fundamental constants of physics.
	\item \emph{Physical interpretation of the normalisation constant.} Even though we argued above that option 2 is not less fundamental than option 1, it remains that in option 1, the normalisation constant $N$ has a physical meaning: it is the frequency, in Hz, of the specific clock transition used as a definition. On the contrary, the normalisation constant $N$ of option 2 does not directly map to the frequency of a particular physical quantity in Hz (beyond the hypothetical physical realisation of the geometric mean~(\ref{eq:geom})). This criticism is certainly valid, but is arguably inessential, because the actual value of the normalisation constant $N$ in both options is accidental (chosen for continuity with historical definitions of the second); and furthermore, the recommended frequencies derived from the defining values of the unit have a physical meaning, and are used in practice to realize the second with option 2.
	\item \emph{Realization uncertainty}. Both options 1 and 2 necessarily yield a realisation uncertainty, given that actual frequency standards always deviate to some extent from the idealized unit. For instance, atoms interact with the ambient black-body radiation, and correcting for this effect involves the imperfect knowledge of the atomic polarisability over the black-body spectrum. As a consequence, the so-called ``Primary Frequency Standards'' (PFS) of the current definition and of option 1 are not genuine PFS because they do not strictly  implement the definition of the second (non-interacting atoms at rest and at 0~K). From this point of view, the fact that the notion of PFS disappears in option 2 has not much physical significance: in practice, all frequency standards must be corrected to match the definition, and the uncertainty on this correction has to be estimated regardless of physical origin.

	A significant difference still remains between options 1 and 2: while the realisation uncertainty solely arise from characteristics of a single atomic species in option 1, it also depends on the knowledge of frequency ratios in option 2. As a consequence, option 2 suffers from an additional realisation uncertainty as compared to option 1. However, three observations mitigate this drawback.

	First, frequency ratios, as measured physical quantities are often more reliable than other physical quantities such as atomic polarisability. Indeed, the latter are usually measured with high precision on dedicated experiments~\cite{PhysRevLett.108.153002, PhysRevLett.109.263004}, and seldom reproduced in different laboratories nor at different times. On the contrary, frequency ratios are usually repeatedly measured, and their consistency is cross-checked by the fitting procedure performed by the WGFS.

	Second, the added realisation uncertainty of option 2 tends to be negligible as the number of species entering the definition increases~\cite{Lodewyck_2019}. This averaging effects comes from the multiplicity of connections between a specific transition and the definition, as illustrated in figure~\ref{fig:connection}.

	Third, although the realisation uncertainty in option 2 is a little increase for the transition that would be chosen as the PFS of option 1, it is more significantly decreased for every transition acting as SRS of option 1. As a result, in a situation where SRS are widely used, as it has been the case for contributions to TAI for the last couple of years, the overall realisation uncertainty could actually be lower with option 2, as compared to option 1.
	\item \emph{Fit industrial need.} In can be argued that industries developing frequency standards could be less confused if a single species is put forward as PFS (option 1), rather than being offered the choice to select one of the several transitions composing the unit of option 2. However, the past experience does not support this argument: driven by technological need, other atoms than Cs have been widely considered in the industry, such as Rb for microwave clocks ranging from small chip scale clocks to large higher performance atomic clocks~\cite{9316270}. In the last few years, collaborations between academic laboratories and industries led to the demonstration of industrial-grade optical clocks~\cite{takamoto2020test, STUHLER2021100264}.
	\item \emph{Compatibility with national legislations}. Because the legal time of most countries is based on UTC, itself based on the SI second, it is necessary that the new definition of the second is compatible with national legislations. While this is most certainly true for option 1, it is less obvious for option 2, especially if allowing for the dynamic updates of defining values, as described in  section~\ref{sec:updates}. For option 2, the compatibility with legislations must therefore be anticipated and verified with national delegates.
	\item \emph{Understanding by the general public}. Before 2018, the second was defined as ``the duration of 9\,192\,631\,770 periods of the radiation corresponding to the transition between the two hyperfine levels of the ground state of the caesium 133 atom''. Although the wording of the definition is now different, this physical interpretation can still be used to explain to the general public how the second is defined, and such an explanation remains valid in the case of option 1. In contrast, option 2 cannot be explained in these terms, as the expression~(\ref{eq:geom}) of the unit is mathematically more involving. However, it is probable that the general public can understand that, given that many different high performing atomic clocks are available, the second is defined from an average (mentioning the weighted geometric mean in unnecessary) of these frequency standards, and that it is practically realizable with any clock, because we have very good measurements of the frequency ratios between these clocks. Indeed, the idea of averaging can be found behind many concepts in other domains, such as the computation of inflation, that the non-specialist public can easily grasp. It can also be argued that such an explanation does not seem more difficult to understand than how fixing the Planck constant can define the kg, especially when it comes to explaining the mise en pratique.
\end{itemize}

The purpose of this section was to focus on the specific advantages and drawbacks of option 2. Making an enlightened choice for the new definition of the second will rather require an analysis of both options on an equal footing. This may prove to be difficult, as different strengths and weaknesses may be hard to weight against each other.

Another strategy could be to start from a list of \emph{a priori} requirements and specifications the new definition should or could meet (\emph{e.g.} ability to reach a consensus, adapted to applications, durability, opportunity to build realisations, uncertainty of realisations, acceptability, \ldots), and then to rate options against these criteria.

\section{Evolution of the unit}
\label{sec:updates}

After the SI second had been defined using the caesium hyperfine transition in 1967, caesium beam clocks, and then caesium fountain clocks have remained the best frequency standards for the following 40 years, only being surpassed by optical frequency standards in the late 2000s.

Ideally, such a situation would be met after the SI second has been redefined, either with a single transition having all the weight (option 1) or with a set of transitions (option 2), this choice remaining relevant throughout the 21st century. However, the field of optical frequency metrology is currently evolving fast, and there is a significant probability that the choice settled in the next few years will become obsolete within a couple of decades. Consequently, before the second is redefined, this eventuality has to be considered, and a clear strategy to face it has to be elaborated. Several possibilities can be considered:
\begin{enumerate}[start=0,label={(\arabic*):}]
	\item The chosen optical second can remain unchanged for more than half a century, disregarding the eventual progress of new frequency standards. Frequency ratios between these standards would be measured with an uncertainty lower than the uncertainty of the best realisations of the SI second.
	\item A new redefinition is adopted a decade or two after the initial redefinition with optical frequency standard(s).
	\item The new definition allows for a regular update of the defining values by the CIPM (International Committee for Weights and Measures), following predefined rules.
\end{enumerate}

All three options above have disadvantages: (\subopzero) means that we are satisfied with a SI second disconnected from the uncertainty of the best clocks, similarly to the current situation with the Cs-based definition. Therefore it would question the motivation to redefine the second based on optical frequency standards at all. (\subopone) means that we may be in a continuous redefinition process, as the interval between redefinitions matches the time required to reach a consensus on a new defintiion. (\suboptwo) implies that the update process becomes partly independent of the CGPM, but more fundamentally that more frequent steps are introduced in the definition of the second, with their associated uncertainty.

It is important to note that these considerations about how the unit may evolve in the near future are essentially independent of whether option 1 or option 2 is chosen: we may exhaustively call these various possibilities options 1.\subopzero, 1.\subopone, 1.\suboptwo, 2.\subopzero, 2.\subopone, and 2.\suboptwo. However, option 2.\suboptwo{} has a specific behaviour when compared to 1.\suboptwo. Indeed, when changing the definition of the second, a discrepancy  is introduced between the old and the new definition, due to our imperfect knowledge of frequency ratios. Quantitatively, the variance of the step in the definition is given, in relative units, by~\cite{Lodewyck_2019}:
\begin{equation}
	\label{eq:vm}
	v_m = \sum_{k,l} (w'_k - w_k)(w'_l - w_l)\Sigma'_{k,l},
\end{equation}
where the double sum runs over all transitions composing the unit, $w'_i$ and $w_i$ are the new and old weights of transition $i$ respectively, and $\Sigma'_{k,l}$ is the output covariance matrix of the global fit of frequency ratios, expressed in relative units\footnote{Using notations of~\cite{Lodewyck_2019} appendix A, we define $\Sigma'_{k,l} = \Sigma_{k,l}/\bar\nu_k\bar\nu_l$, such that the relative uncertainty on the frequency ratios $\rho_{k,l}$ is given by $(\delta \rho_{k,l}/\rho_{k,l})^2 = \Sigma'_{k,k} + \Sigma'_{l,l} - 2 \Sigma'_{k,l}$}. In the specific case of option 1, where changing from transition $i$ to transition $j$, the expression~(\ref{eq:vm}) reduced as expected to:
\begin{equation}
	v_m = \left(\frac{\delta \rho_{i,j}}{\rho_{i,j}}\right)^2 \quad \textrm{for option 1},
\end{equation}
\emph{i.e.} the typical step in the unit is given by the relative uncertainty of the frequency ratio between the new and the old definition.

After repeatedly changing the definition, the steps $v_m$ accumulate, resulting in a random walk of the unit. From equation~(\ref{eq:vm}), we can see that option 2.\suboptwo{} has a clear advantage over options 1.\subopone{}, 1.\suboptwo{} and 2.\subopone{}: a regular but smooth update of the weights yields smaller steps in the redefinition, mitigating the risk of divergence of the unit. In this process,  the definition should be changed only if $v_m$, which is is an objective and quantitative criterion which can be calculated from the global fit of frequency ratios, significantly reduces from one redefinition to the next. The main drawback of option 2.\suboptwo{} is that it constitutes a more radical change from the traditional way the units are defined.

Options 1.\subopone{}, 1.\suboptwo{}, 2.\subopone{}, and 2.\suboptwo{}, imply that the unit will change frequently, which could be a concern if the second eventually becomes incompatible with an anterior definition, after several steps have accumulated. This is all the more a concern that all other units of the SI (except the mole) are derived from the second, although in practice these other units are realized with uncertainties several orders of magnitude larger than for the second. In fact, it can be shown that the random walk resulting from the successive definition is actually convergent if the series of $v_m$ is convergent. In this case, the well-defined asymptotic value of the definition can be seen as the actual, immovable definition of the second; And at any time, the definition is an approximation of this asymptotic value, with a discrepancy that is guaranteed to be smaller than the uncertainty of the best realizations of the time.

\section{Conclusion}

In this paper, we discussed the possibility to redefine the second using the weighted geometric mean of several clock transitions, as a generalisation of the definition based a single species promoted as a primary frequency standard. This proposition, more able to reach a consensus given the diversity of optical frequency standards under development, is inspired by, and makes use of the decade-long work of the WGFS in collecting frequency ratio measurements and in compiling them into a set of consistent best estimates. From this point of view, option 2 aims at taking advantage, in the definition, of this effort: A proper choice of weights allows to realize the second with a vast variety of frequency standards, while minimizing the overall realisation uncertainty.

Because it deviates more from the current definition of the second, option 2 legitimately raises concerns and questions, such as its relation to fundamental constants, the physical signification of the normalisation constant $N$, or the physical origin of its realisation uncertainty. We have proposed answers to these questions, starting from an introspection about how these concepts are dealt with in the current definition or in option 1, and highlighting the many similarities between the options.

How the new definition may evolve in the next decades is a concern that should be addressed, regardless of whether option 1 or 2 is selected. If optical frequency standards continue to be improved, all foreseeable scenarios have drawbacks. However, assuming that the definition will be eventually be updated in order to follow the progress of frequency standards, option 2 offers a smoother transition with a reduced gap, owing to its ability to continuously tune the weights of different species, as opposed to option 1 for which the weights discretely change from one species to another. Therefore, we could take advantage of this property, and allow for regular updates of the unit, based on quantitative and predefined criteria.

\section*{Acknowledgments}

We acknowledge funding from the French ``Agence Nationale de la Recherche'' (ANR), project RYCARD (ANR-22-CE47-0009, 2022)

\section*{References}

\bibliographystyle{iopart-num}
\bibliography{SFSM_proceedings}

\end{document}